# Interference effect on Raman spectrum of graphene on $SiO_2$/Si

Duhee Yoon,[1] Hyerim Moon,[1] Young-Woo Son,[2*] Jin Sik Choi,[3] Bae Ho Park,[3] Young Hun Cha,[4] Young Dong Kim,[4] and Hyeonsik Cheong[1§]

*[1]Department of Physics, Sogang University, Seoul 121-742, Korea,*

*[2]School of Computational Sciences, Korea Institute for Advanced Study, Seoul 130-722, Korea,*

*[3]Division of Quantum Phases & Devices, School of Physics, Konkuk University, Seoul 143-701, Korea,*

*[4]Department of Physics, Kyung Hee University, Seoul 130-701, Korea,*

**Abstract**

The intensity ratio between two major Raman bands in graphene is one of the most important information for physics of graphene and has been believed to represent various intrinsic properties of graphene without critical assessment of extrinsic effects. We report a micro Raman spectroscopy study on the Raman intensity ratio of the 2*D* band to the *G* Raman band of graphene varying the thickness of dielectric layers ($SiO_2$) underneath it. The ratio is shown to change by almost 370% when the thickness is varied by 60%. The large variation in the ratio is well explained by theoretical calculations considering multiple Raman scattering events at the interfaces. Our analysis shows that the interference effect is critical in extracting the intrinsic 2*D* to *G* intensity ratio and therefore must be taken into account in extracting various physical properties of graphene from Raman measurements.

## I. INTRODUCTION

Graphene, a two-dimensional hexagonal crystal of carbon atoms, has attracted immense interests from researchers in various disciplines because of its novel electronic properties, such as a high carrier mobility[1-3] and anomalous quantum Hall effect.[4,5] These intriguing properties are caused by the linear energy dispersion versus momentum around the Dirac points.[1-5] After the first successful isolation of graphene,[1] its unique physical properties have been studied by using various experimental tools: two or four terminal transport measurement,[4,5] Raman spectroscopy,[6] spin transport,[7] infrared spectroscopy,[8] angle resolved photoemission spectroscopy,[9] and scanning tunneling microscopy.[10,11]

Among these experimental probes, Raman spectroscopy is one of the most successful tools in investigating the electronic and structural properties of graphene.[6,12-22] A typical Raman spectrum of graphene consists of two major features (Fig. 1): the $G$ ($\omega_G \sim 1586$ cm$^{-1}$) and the $2D$ ($\omega_{2D} \sim 2686$ cm$^{-1}$) bands.[6,12,13] The $G$ band originates from the Stokes Raman scattering with one phonon ($E_{2g}$) emission.[23] It is known that as the doping concentration is increased, its frequency blueshifts and its width decreases.[16-21] The $2D$ band (or sometimes called $G'$) is due to the Stokes-Stokes double resonant Raman scattering with two-phonon ($A_1'$) emissions,[6,15] and its shape is very sensitive to the number of graphene layers.[6,12-14] Although the absolute intensity of either of these peaks in a Raman spectra may depend on various external factors such as equipment alignment that may vary in each measurement, the intensity ratio of the $2D$ band to the $G$ band ($I_{2D}/I_G$) is often thought to be immune to such external factors and represent the intrinsic properties of a given type of graphene. It has been used to determine basic structural and electronic properties of graphene such as the number of layers,[6,12-14] doping concentration,[16-21] and optical anisotropy.[22]

In this Letter, we find that through micro-Raman spectroscopy,[14,22] the observed ratio $I_{2D}/I_G$ varies by 370 % when the thickness of the $SiO_2$ layer on silicon substrates is varied by 60 %. Hence, the thickness of the $SiO_2$ layer, which should not affect the intrinsic properties of the graphene sample on it, should be taken into account in interpreting the observed $I_{2D}/I_G$. By considering the interference of the excitation laser as well as the Raman signal due to multiple reflections at the interfaces, one can calculate the enhancement factors for the $2D$ and the $G$ bands. We show that these factors are different in general, and therefore, one needs to factor out these enhancement factors properly when important intrinsic properties such as $I_{2D}/I_G$ are deduced from the experimental data.

## II. EXPERIMENTAL

Mechanically exfoliated graphene samples[1] were placed on top of $SiO_2$/Si substrates with various $SiO_2$ layer thicknesses (~240 to ~380 nm). Initially, Si (p-type) substrates covered with ~300- or ~388-nm $SiO_2$ layer were prepared by wet thermal oxidation. Then, the thickness of the $SiO_2$ layer was reduced by wet etching in a buffered $NH_4F$-HF (BHF) solution for various etching times. The surface roughness of unetched $SiO_2$ is about 0.16 nm as measured by atomic force microcopy. Etching increases the roughness, but all the etched substrates have more or less similar roughness (0.54–0.62 nm).

Since the thickness and refractive index of the $SiO_2$ layer are crucial factors in the analysis of the enhancement factors for the Raman intensity, we used high-precision spectroscopic ellipsometry (SE) to determine their precise values. The SE measurement[24,25] was performed under high purity $N_2$ atmosphere in the wavelength range from 190 to 1100 nm. To reduce experimental errors we measured the SE spectra at multiple incidence angles of 60, 70, and 80°, and then extracted a single set of parameters to fit all 3 spectra.[26] The error bars in the determination of the thickness and the refractive index are ±0.3–2.0 nm and $2.0\times10^{-4}$, respectively.

Single layer graphene samples are roughly identified with an optical microscope, and then confirmed by micro-Raman spectroscopy measurements. Because of the interference effect between the graphene and $SiO_2$ layers, the color and the contrast of graphene are influenced by the wavelength of the illumination and the thickness of the $SiO_2$ layer.[27-30] In general, it is known that a ~300-nm thick $SiO_2$ layer is optimal for white light illumination.[27] Since graphene samples on $SiO_2$ layers thicker than 340 nm are not visible under white light illumination, a red dichroic filter was used for these samples (Fig. 1(a) and (b)).

For the micro Raman measurements, the 514.5-nm (2.41 eV) line of an Ar ion laser was used as the excitation source, and the laser power was ~1 mW. The laser beam was focused onto the graphene sample by a 40× microscope objective lens (0.6 N.A.), and the scattered light was collected and collimated by the same objective. The scattered signal was dispersed with a Jobin-Yvon Triax 550 spectrometer (1200 grooves/mm) and detected with a liquid-nitrogen-cooled CCD detector. The spatial resolution was less than 1 μm, and the spectral resolution was about 1 $cm^{-1}$. The single layer graphene was exactly identified by the unique shape of the Raman 2*D* band as shown in Fig. 1.[6,12-14]

## III. RESULTS AND DISCUSSION

We found that the Raman intensities of the *G* and 2*D* bands are indeed strong functions of the thickness of $SiO_2$ layer [Fig. 2(a)]. As the thickness is increased from 240 nm, the observed intensities increase first and show the highest values at the thickness of ~280 nm for the *G* band and ~290 nm for the 2*D* bands, respectively. Since the two maxima occurs at different wavelengths, the resulting observed $I_{2D}/I_G$ ratio varies greatly; the maximum of the intensity ratio is around 9.3 and 6 times higher than the minimum (Fig. 2(b)). Some scatter in the experimental data will be discussed later.

To explain the observed peculiar variation of the intensity, we use the multi-reflection model (MRM) of the Raman scattered light (Fig. 3 (a) and (b)). In this model, the absorption and scattering processes are treated separately. We note that a similar method[31] was applied to explain the variation of the *G* band intensity when the number of graphene layers increases. We also note that it was applied schematically to Raman intensity variation as a function of the thickness of the dielectric layer without considering the difference in the wavelengths of the laser and Raman scattered light.[31] In general, the wavelengths of the Stokes Raman scattered light and the laser are taken to be the same in similar calculations.[31-34] However, in graphene, the actual differences between the wavelengths of the laser, the Raman *G* band, and the 2*D* band are quite large. When the 514.5-nm (2.41 eV) line of an Ar ion laser is used, the Raman *G* and 2*D* bands of single layer graphene are located at ~1586 $cm^{-1}$ and ~2686 $cm^{-1}$, respectively. In terms of wavelengths, these correspond to 560.2 nm (2.21 eV) and 597.0 nm (2.08 eV), respectively. Hence, these differences and concomitant differences in the index of refraction will result in different interference patterns for the Raman bands.

In the MRM, as shown in Fig. 3(a), the laser beam is absorbed by the π-electrons of graphene while passing through the graphene layer. However, the laser beam goes through multiple reflections inside the graphene layer as well as in the $SiO_2$ layer. Due to these multiple reflections, there are multiple chances for the beam to be absorbed by the π-electrons. The net absorption term ($F_{ab}$) could be represented by the sum of the dots in Fig. 3(a) and can be expressed as

$$F_{ab} = t_1 \frac{\left[1 + r_2 r_3 e^{-2i\beta_2}\right] e^{-i\beta_x} + \left[r_2 + r_3 e^{-2i\beta_2}\right] e^{-i(2\beta_1 - \beta_x)}}{1 + r_2 r_3 e^{-2i\beta_2} + (r_2 + r_3 e^{-2i\beta_2}) r_1 e^{-2i\beta_1}}, \quad (1)$$

where $t_1 = 2n_0 / (\tilde{n}_1 + n_0)$ , $r_1 = (n_0 - \tilde{n}_1) / (n_0 + \tilde{n}_1)$ , $r_2 = (\tilde{n}_1 - \tilde{n}_2) / (\tilde{n}_1 + \tilde{n}_2)$ and $r_3 = (\tilde{n}_2 - \tilde{n}_3) / (\tilde{n}_2 + \tilde{n}_3)$ are the Fresnel transmittance and reflection coefficients for the interfaces involving air (0), graphene (1), $SiO_2$ (2), and Si (3). $n_0 = 1$ is the refractive index of air, and $\tilde{n}_1$, $\tilde{n}_2$ and $\tilde{n}_3$ are the refractive indices for graphene, $SiO_2$, and Si, respectively. We also use abbreviations $\beta_x = 2\pi x \tilde{n}_1 / \lambda$, $\beta_1 = 2\pi d_1 \tilde{n}_1 / \lambda$, $\beta_2 = 2\pi d_2 \tilde{n}_2 / \lambda$, where $x$ is the depth of the point where the interaction occurs, and $d_1$ and $d_2$ are the thickness of the single layer graphene and the $SiO_2$ layer, respectively.[35]

Similarly, the net scattering term ($F_{sc}$) could be represented by the sum of the arrow lines in Fig. 3(b) and expressed as

$$F_{sc} = t_1' \frac{\left[1 + r_2 r_3 e^{-2i\beta_2}\right] e^{-i\beta_x} + \left[r_2 + r_3 e^{-2i\beta_2}\right] e^{-i(2\beta_1 - \beta_x)}}{1 + r_2 r_3 e^{-2i\beta_2} + (r_2 + r_3 e^{-2i\beta_2}) r_1 e^{-2i\beta_1}}, \qquad (2)$$

where $t_1' = 2\tilde{n}_1 / (\tilde{n}_1 + n_0)$. $\lambda$ is the wavelength of the excitation source in the net absorption term and is the wavelength of the *G* or 2*D* bands in the net scattering term. Then, the total enhancement factor ($F$) is given by

$$F = N \int_0^{d_1} \left| F_{ab} F_{sc} \right|^2 dx, \qquad (3)$$

where $N$ is a normalization factor, which is a reciprocal number of the total enhancement factor for a free-standing graphene, obtained by replacing the $SiO_2$ and Si layers with air. The measured Raman intensity ($I$) is $I = I_i \cdot F$, where $I_i$ is the intrinsic Raman intensity playing the role of a single fitting parameter in the subsequent calculations.

Figure 3(c) is the calculated enhancement factor ($F$), relative to the case of a free-standing graphene film. The thickness of graphene ($d_1$) is taken to be 0.34 nm, which corresponds to the interlayer distance in graphite crystals. The interference effects on the Raman *G* and 2*D* bands are clearly seen; as a function of the thickness of the $SiO_2$ layer, the enhancement factor for the Raman $G$ and $2D$ bands vary by a factor of up to 48. In Fig. 2(a) and (b), the results from our model calculation are compared with the experimental data for the $I_G$, $I_{2D}$, and $I_{2D}/I_G$. Overall, they show good agreement with each other. We can fit the data in Fig. 2(a) by setting the intrinsic Raman intensity ratio $I_{i,2D}/\ I_{i,G}$ =3.4. Although there is some scatter, the measured data are well represented by the curves fitted from the model calculation. It should be noted that the calculated enhancement without considering the difference in the wavelengths for each Raman scattering event (solid curve in Fig. 3(c)) deviates significantly from the correct ones when the dielectric layer becomes thicker than 200 nm.

We found that some of the scatter in the data for the intensity ratio originate from other extrinsic factors such as defects and doping.[18,19] Some of the samples showed the defect-induced Raman $D$ band (~1350 cm$^{-1}$) signals. These samples tend to have higher $G$ band intensities as compared with those that do not show the $D$ band. In Fig. 2(b), the data from these samples (star symbols) are off the main tendency. Other samples showed higher levels of doping, as indicated by blueshifted $G$ band peak positions and decreased widths of the $G$ band.[16-21] Highly doped samples are known to exhibit lower $I_{2D}/I_G$ ratios.[18,19] The data from the samples with estimated doping densities[16-18,20] in excess of $5\times10^{12}\ \mathrm{cm}^{-2}$ are identified by open circles in Fig. 2(b). By excluding data from such samples, we fit the data from remaining 'intrinsic' samples to obtain $I_{i,2D}/\ I_{i,G}$=3.2.

In the above analysis, we used a value of 0.34 nm as the thickness of single layer graphene. However, the thickness of graphene, which is just one atomic-layer thick, is not a well-defined quantity. AFM measurements[1,12] do not give an definitive answer as to the exact value of the 'thickness' of graphene. Also it is known that there exist ripples on the order of about 0.5 nm, which also affect the 'optical thickness' of the graphene layer. In order to test the sensitivity of our model to the choice of the thickness value, we repeated the calculation while varying the thickness of single layer graphene from 0.1 to 1 nm. However, this changed the interference pattern only slightly: the maxima of the enhancement factors (Fig. 3(c)) shifted by less than 2 nm. There are also speculations that there exists a thin layer of air or water between graphene and $SiO_2$, which might also affect the interference. Again, we calculated the interference pattern as a function of the thickness of interlayer (air or water) in the range of 0–1 nm, which shifted the maxima of the enhancement factors by less than 2 nm. The index of refraction of graphene is another ambiguous quantity. We used the index of refraction of graphite ($\tilde{n}_1$) as a first approximation as other authors did.[27-31] This somewhat arbitrary choice does not affect our result, though, because varying the index of refraction from $0.5\tilde{n}_1$ to $2.0\tilde{n}_1$ changed only the absolute amplitude of the interference pattern but did not shift the interference pattern. As a matter of fact, the interference pattern changed appreciably only when the thickness or the refractive index of $SiO_2$ was varied.

In general, the interference effect is a function of the incident angle of the light. In our analysis, we assumed normal incidence because most micro Raman scattering measurements are performed in backscattering geometry. Even when the numerical aperture (N.A.) of the objective lens is large, the laser beam is almost normally incident on the sample provided that the beam is Gaussian and the focused laser beam hits the sample surface at the beam waist. However, in practice, since the focal depth is about 1 μm for N.A.=0.6 and $\lambda = 500$ nm, it is conceivable that a significant portion of the beam enter the sample at an oblique angle if the focus is only slightly off. We considered the upper bound of the effect of the large N.A. on the interference pattern, regarding a Gaussian distribution of the incident light intensity but treating the beam path with classical ray optics. Contributions from each portion of the beam with an incident angle $\theta$ ($0 < \theta < \theta_{max} = \arcsin \text{N.A.}$) were calculated separately and then integrated over $\theta$.[39] As shown in Fig 4(a) and (b), the patterns of the enhancement factors of the $G$ and $2D$ bands shift slightly for larger N.A. values. The peak positions for 0.9 N.A. is shifted by about 10 nm with respect to those for normal incidence. The ratio of the enhancement factors ($F_{2D}/F_G$) is also slightly shifted in Fig 4(c). New fitting curves including the effect of a large N.A. value of 0.6 are shown in Fig. 2 as dashed curves. These new fitting curves give $I_{i,2D}/ I_{i,G}$=3.5.

We also calculated the enhancement factors of the $G$ ($F_G$) and $2D$ ($F_{2D}$) bands as functions of the thickness of $SiO_2$ layer and the wavelength of the excitation source for normal incidence. The results are shown in Fig. 5(a) and (b). In this calculation, $\omega_G$ is fixed since it does not vary with the laser wavelength. However, the frequency of the Raman $2D$ band ( $\omega_{2D}$ ) depends on the laser wavelength,[37] and the dispersion of the Raman $2D$ band can be given as a linear function, $\omega_{2D} = 2444.24 + 99.06 E_{Laser} \mathrm{cm}^{-1}$ , where $E_{Laser}$ is the laser energy in eV.[38] Figures 5(a) and (b) clearly show that the Raman signal is significantly enhanced or suppressed, depending on the laser wavelength and the $SiO_2$ layer thickness. Figure 5(c) is a contour plot of the ratio of the enhancement factors for the Raman $2D$ band to the $G$ band. It is clear that these factors play a major role in determining the intensity ratio. Our calculation indicate that it is important to factor out the interference effect first, when comparing $I_{2D}/I_G$ data from samples with different $SiO_2$ layer thicknesses, or obtained with different lasers. It should also be noted that this kind of interference effect is not unique to Raman measurements but applies to any spectroscopic measurements on thin samples on dielectric layers. Appropriate choice of the dielectric layer thickness therefore can significantly enhance the measured signal in such cases.

## IV. SUMMARY

In summary, a strong dependence of the Raman spectrum of single layer graphene on the thickness of the $SiO_2$ layer on the substrate is observed and analyzed in terms of multiple reflection interference. It is found that the Raman spectrum depends not only on the $SiO_2$ layer thickness but also on the wavelength of the excitation laser. This effect significantly influences the observed intensity ratio of the Raman 2*D* band to the *G* band.

**ACKNOWLEDGMENT**

This research was supported by Basic Science Research Program through the National Research Foundation of Korea (NRF) funded by the Ministry of Education, Science and Technology (MEST) (R01-2008-000-10685-0 and KRF-2008-314-C00111). D. Y. was supported by the Seoul Science Fellowship. Y.-W. S. was supported in part by the NRF grant funded by MEST (Quantum Metamaterials Research Center, R11-2008-053-01002-0 and Nano R&D program 2008-03670). B.H.P. was partly supported by WCU program through the NRF funded by MEST (R31-2008-000-10057-0). The work at KHU was supported by the National Research Laboratory Fund through Nano Optical Property Laboratory.

## REFERENCE

*hand@kias.re.kr

§hcheong@sogang.ac.kr

# CAPTIONS

**FIG. 1**. (Color online) Raman spectrum of a single layer graphene sample taken with a 514.5-nm laser as the excitation source. Insets (a) and (b) are the optical microscope images of a single layer graphene sample on a 350-nm $SiO_2$ layer with white and red (filtered, 615-730 nm) light illumination, respectively. The graphene layer can be easily seen in (b), but not in (a).

**FIG. 2**. (Color online) (a) *G* (circle dots) and 2*D* (square dots) band Raman intensities as functions of the thickness of the $SiO_2$ layer. (b) Raman intensity ratio $I_{2D}/I_G$ as a function of the thickness of the $SiO_2$ layer. The stars represent data taken from Raman spectra which showed the *D* band (~1350 $cm^{-1}$), and the open circles represent data taken from the samples with high doping. The curves in (a) and (b) are the calculation results based on the MRM model. The inset is the calculated result for 0 to 500 nm. The dashed curves are the results when the effect of the large N.A. is included for N.A.=0.6.

**FIG. 3.** (Color online) Schematic diagrams of multiple reflection interference in the (a) absorption and (b) scattering processes. $n_0$, $\tilde{n}_1$, $\tilde{n}_2$ and $\tilde{n}_3$ are the refractive indices of air, graphene, $SiO_2$, and Si, respectively. $d_g$ and $d_{SiO_2}$ are the thickness of graphene and $SiO_2$ layer, respectively, and *x* is the depth in the graphene layer. The dots are the points of interaction between the laser beam and the π-electrons of graphene. (c) Calculated Raman intensities as a function of the thickness of the $SiO_2$ layer. The solid curve was obtained in a simple approximation where the Raman bands and the laser are taken to have the same wavelength.

**FIG. 4.** (Color online) Calculated Raman enhancement factors of (a) *G* band and (b) 2*D* band as functions of the thickness of the $SiO_2$ layer for various values of the numerical aperture of the objective lens. The excitation wavelength is taken to be 514.5 nm. (c) Ratio of the enhancement factor for the 2*D* band to that of the *G* band, $F_{2D}/F_G$.

**FIG. 5.** (Color online) Plots of calculated Raman enhancement factors of (a) *G* band and (b) 2*D* band as functions of the thickness of $SiO_2$ layer and the wavelength of the laser. (c) Ratio of the enhancement factor for the 2*D* band to that of the *G* band, $F_{2D}/F_G$.

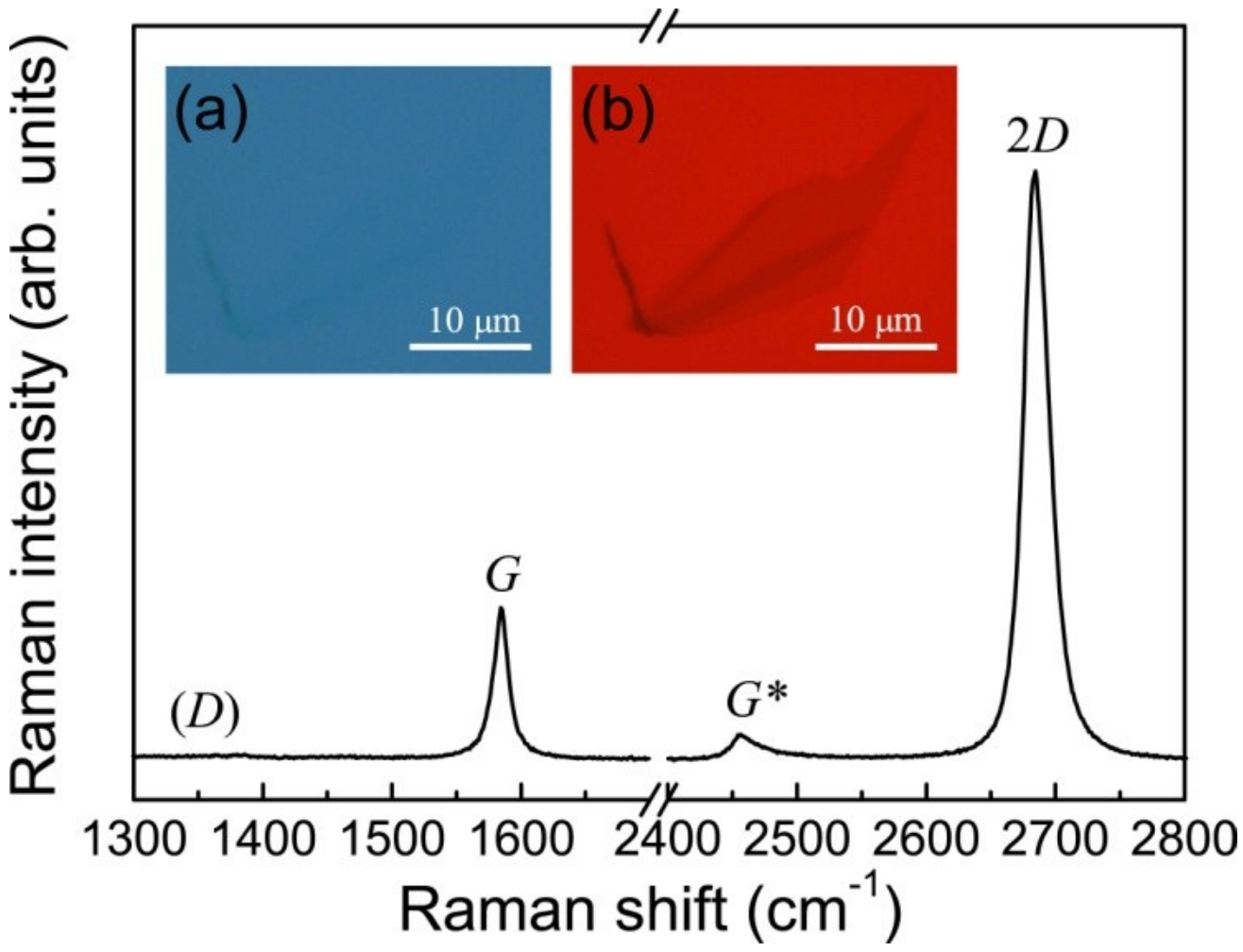
(a)
10 μm
(b)
10 μm
2D
G
G*
(D)
Raman intensity (arb. units)
1300
1400
1500
1600
2400
2500
2600
2700
2800
Raman shift (cm-1)

**FIGURE 1**

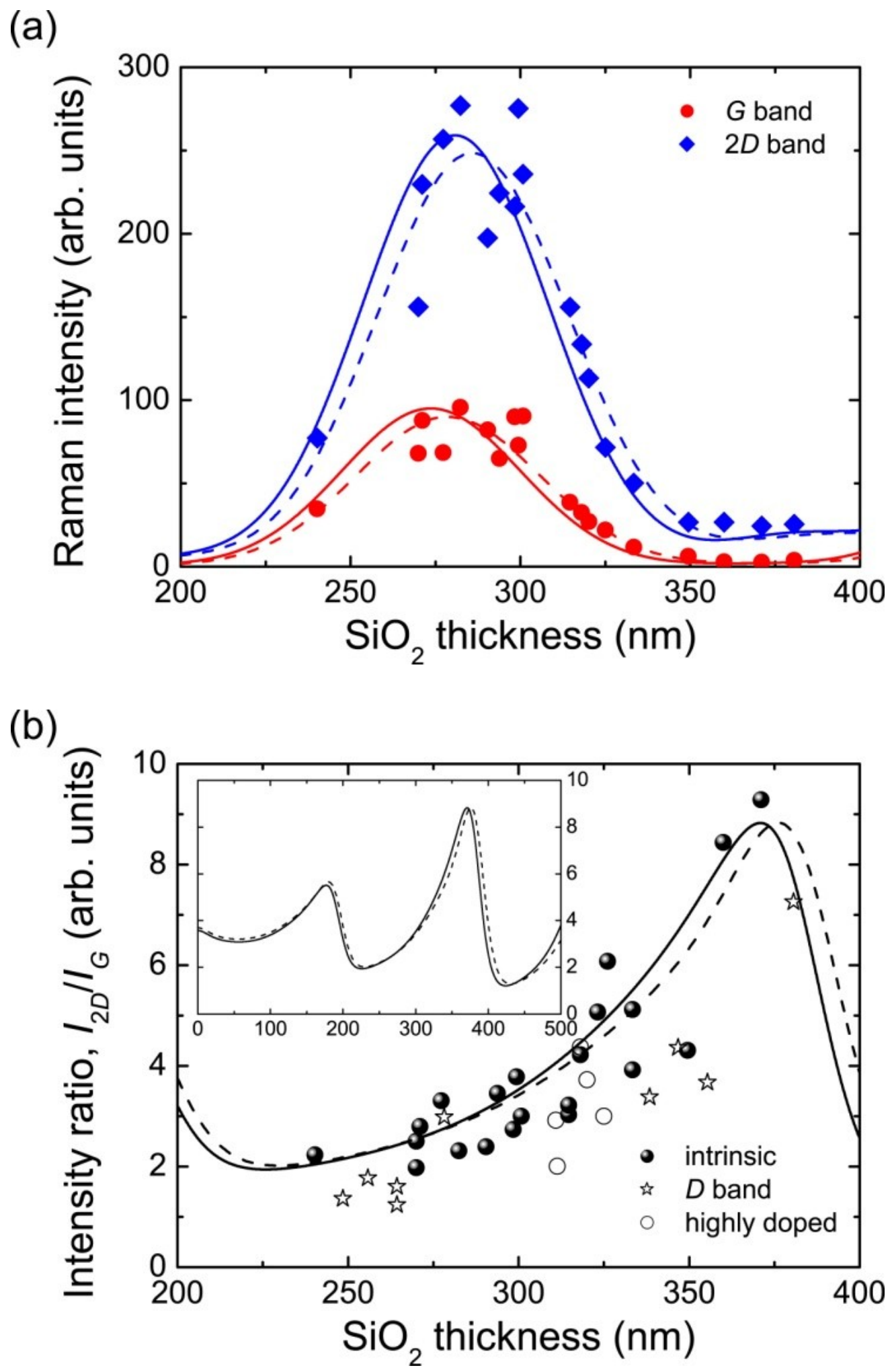
(a)
Raman intensity (arb. units)
G band
2D band
SiO2 thickness (nm)
(b)
Intensity ratio, I2D/IG (arb. units)
intrinsic
D band
highly doped
SiO2 thickness (nm)

**FIGURE 2**

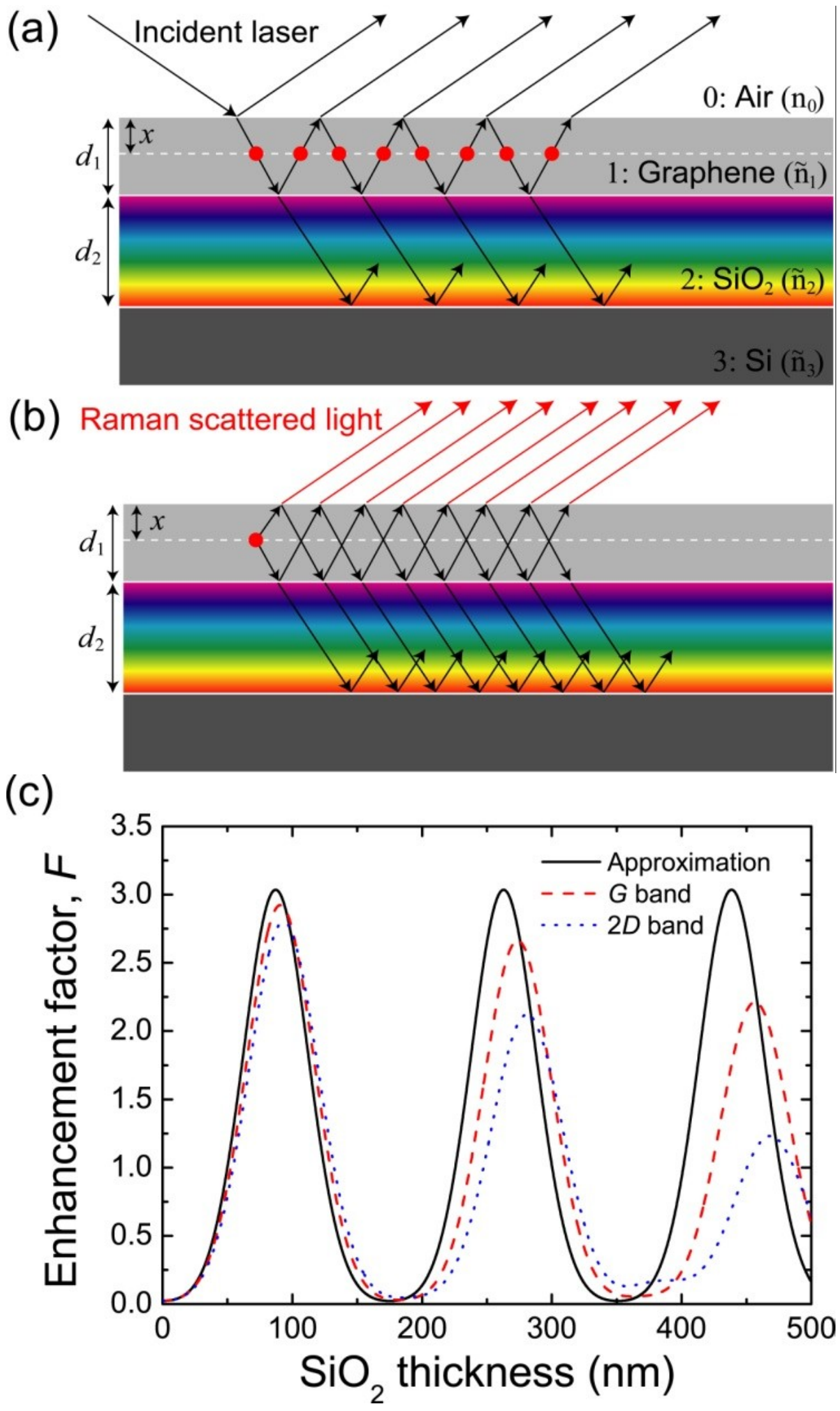
(a)
Incident laser
0: Air ($n_0$)
$x$
$d_1$
1: Graphene ($\tilde{n}_1$)
$d_2$
2: $SiO_2$ ($\tilde{n}_2$)
3: Si ($\tilde{n}_3$)
(b)
Raman scattered light
$x$
$d_1$
$d_2$
(c)
Enhancement factor, $F$
$SiO_2$ thickness (nm)
Approximation
$G$ band
$2D$ band
0.0
0.5
1.0
1.5
2.0
2.5
3.0
3.5
0
100
200
300
400
500

**FIGURE 3**

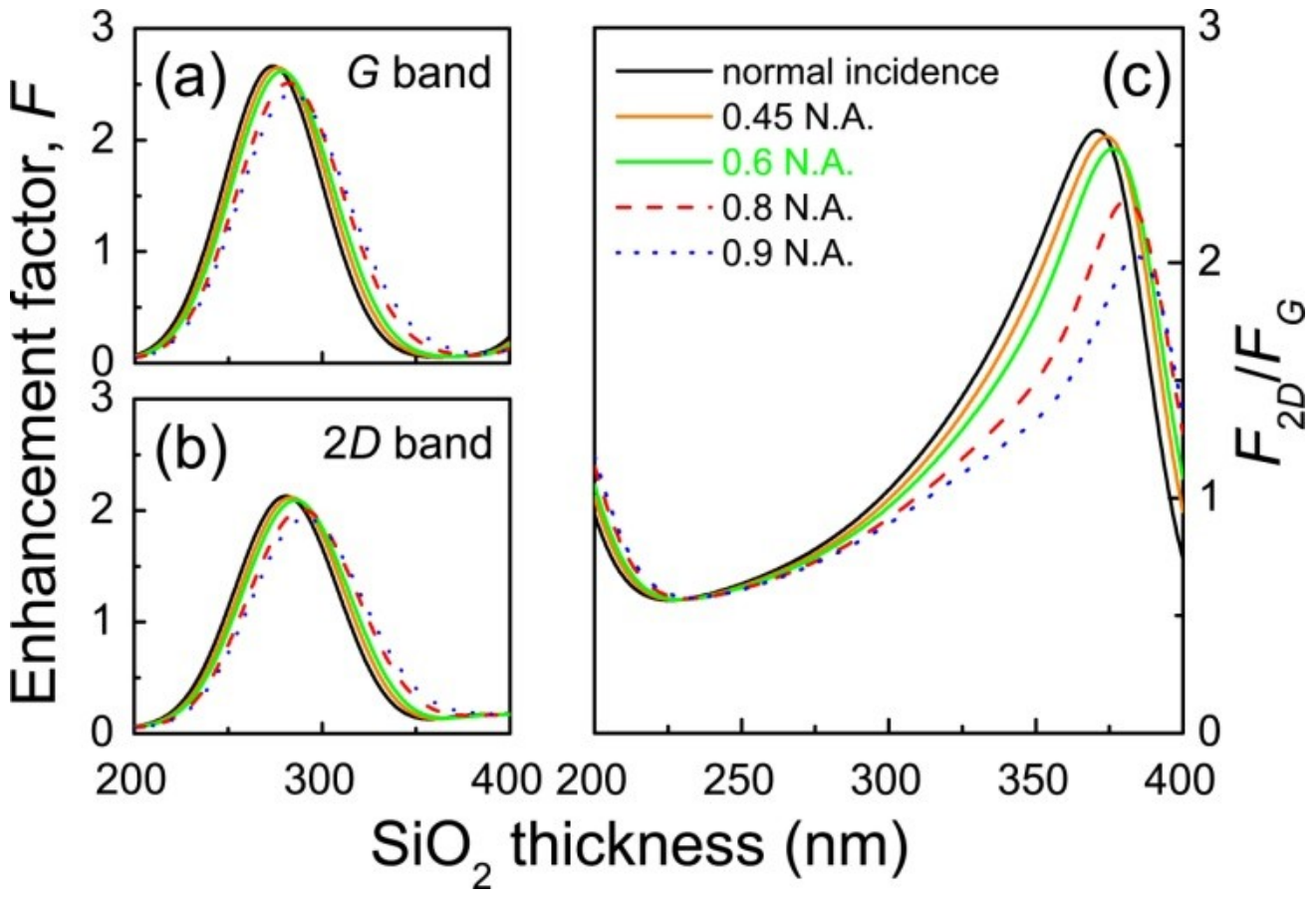

(a)
G band
(b)
2D band
(c)
normal incidence
0.45 N.A.
0.6 N.A.
0.8 N.A.
0.9 N.A.
Enhancement factor, F
$F_{2D}/F_G$
SiO$_2$ thickness (nm)

**FIGURE 4**

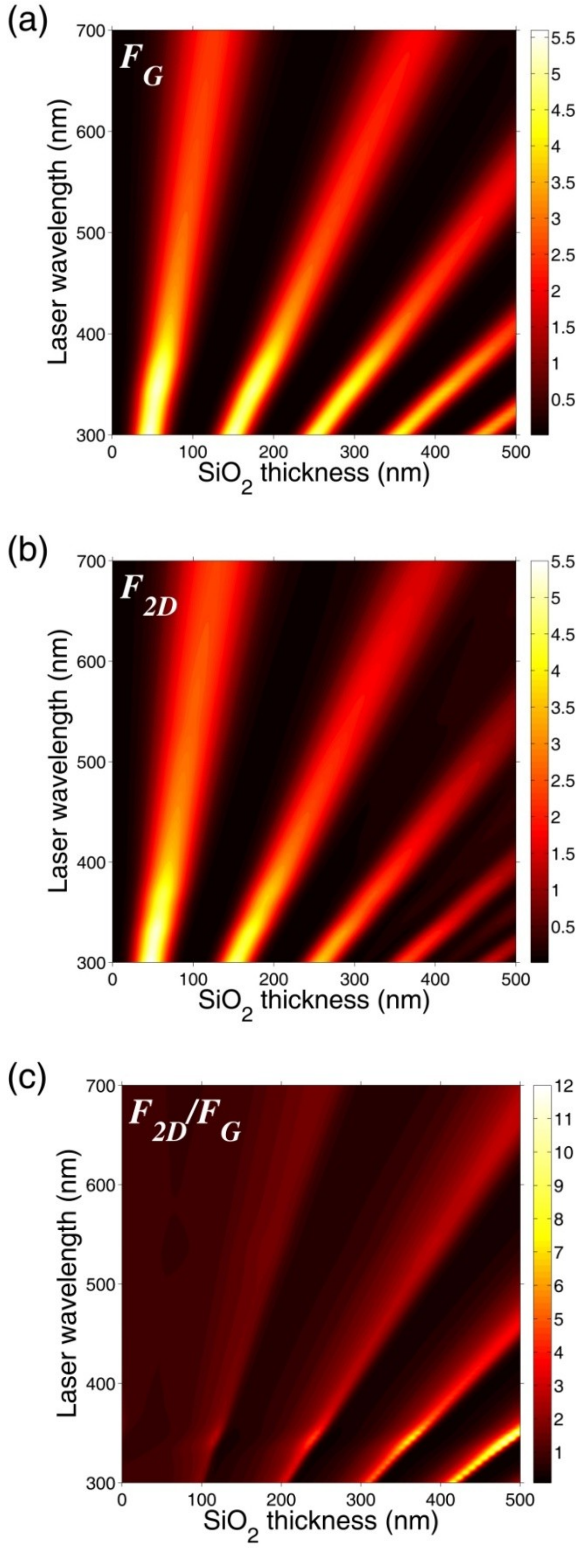
(a)
$F_G$
Laser wavelength (nm)
$SiO_2$ thickness (nm)
(b)
$F_{2D}$
Laser wavelength (nm)
$SiO_2$ thickness (nm)
(c)
$F_{2D}/F_G$
Laser wavelength (nm)
$SiO_2$ thickness (nm)

**FIGURE 5**